\documentclass{appolb}
\usepackage{graphicx}

\usepackage{amsmath}
\usepackage{dcolumn}
\usepackage{bm}

\usepackage{amssymb}
\usepackage{caption,subcaption}
\usepackage{dsfont}
\usepackage{url}

\def\pp{\mathbf{p}}

\def\lan{\left\langle}
\def\ran{\right\rangle}
\def\delpp{\Delta^{\alpha \beta} p_{\alpha}p_{\beta}}

\begin{document}

\title{Novel relaxation time approximation: a consistent calculation of transport coefficients with QCD-inspired relaxation times%
\thanks{Presented at the XXIXth International Conference on Ultrarelativistic Nucleus-Nucleus Collisions (QUARK MATTER 2022)}%
}
\author{Gabriel S.~Rocha, Gabriel S.~Denicol
\address{Instituto de F\'{\i}sica, Universidade Federal Fluminense, Niter\'{o}i, Rio de Janeiro,
Brazil}
\\[3mm]
{Maurício N. Ferreira 
\address{Instituto de Física Gleb Wataghin, Universidade Estadual de Campinas, Campinas, São Paulo, Brazil}
}
\\[3mm]
Jorge Noronha
\address{Illinois Center for Advanced Studies of the Universe \& Department of Physics, University of Illinois Urbana-Champaign, Urbana, IL 61801, USA}
}
\maketitle
%
\begin{abstract}
We use a novel formulation of the relaxation time approximation to consistently calculate the bulk and shear viscosity coefficients using QCD-inspired energy-dependent relaxation times and phenomenological thermal masses obtained from fits to lattice QCD thermodynamics. The matching conditions are conveniently chosen to simplify the computations.
\end{abstract}
  
\section{Introduction}

Nuclear matter in extreme conditions can be investigated through ultra-relativistic heavy-ion collisions. In particular, obtaining the transport coefficients of the quark-gluon plasma, throughout the QCD phase diagram, is a very challenging task that is currently beyond the reach of first-principles techniques \cite{Meyer:2011gj}.
In this contribution, we compute the transport coefficients of an effective kinetic model \cite{Alqahtani:2015qja,Rocha:2022fqz} with a temperature-dependent mass whose equation of state mimics lattice QCD thermodynamics \cite{Borsanyi:2010cj}. We use the new relaxation time approximation (RTA) of the relativistic Boltzmann equation proposed in \cite{Rocha:2021zcw} and impose alternative matching conditions such that the interaction energy \cite{Jeon:1995zm} depends only on the temperature even out of equilibrium.

\section{The quasi-particle model}

The relativistic Boltzmann equation for quasi-particles with a temperature-dependent mass, $M(T)$, is given by \cite{Denicol:2021},
\begin{equation}
\label{eq:BoltzmannQuasi}
p^\mu \partial_\mu f_{\bf p} + \frac{1}{2}\partial_i M^2(T) \partial^i_{(\bf p)} f_{\bf p} = C\left[ f_{\bf p}\right],
\end{equation}
where $f_{\bf p} = f(x,\bf{p})$ is the single particle distribution function. Above, $\partial^i_{(\bf p)}  = \partial/\partial \mathbf{p}_{i}$, and $C\left[ f_{\bf p}\right]$ is the collision integral.

In the limit of vanishing net-charge, the main dynamical equation is the continuity equation for the energy-momentum tensor, $T^{\mu\nu}$,
\begin{equation}
\label{eq:Tmunucons}
\partial_\mu T^{\mu\nu} = 0.
\end{equation}
In the presence of a thermal mass, 
$T^{\mu\nu} \equiv \langle p^\mu p^\nu \rangle  + g^{\mu\nu} B$, where $B$ is the interaction energy \cite{Jeon:1995zm}, $g^{\mu \nu}$ denotes the metric, $\langle \cdots \rangle = \int dP \cdots f_{\bf p}$,  $\int dP = g \int d^3 \mathbf{p}/[(2\pi)^3 E_{\bf p}]$, with $g$ being the degeneracy factor and $E_{\bf{p}} = \sqrt{ \mathbf{p}^2 + M^2 }$. The interaction energy $B$ satisfies the following dynamical equation, 
\begin{equation}
\label{eq:dB}
\partial_{\mu} B = - \frac{1}{2} \partial_{\mu} M^2 \langle 1 \rangle,
\end{equation}
which is valid both in and out of equilibrium. We consider Maxwell-Boltzmann statistics so that in equilibrium $f_{\bf{p}} = \exp\left( - \beta u_{\mu}p^{\mu} \right) \equiv f_{0\bf{p}}$, with $\beta=1/T$ and $u_{\mu}$ being the fluid 4-velocity (which satisfies $u_\mu u^\mu = 1$).

The temperature dependence of the mass is obtained such that the equation of state of the model describes lattice QCD results \cite{Borsanyi:2010cj}.
Plots for $B(T)$ and $M(T)$ can be seen in Refs.~\cite{Alqahtani:2015qja,Rocha:2022fqz}. Qualitatively, $M(T)/T$ is very large at low temperatures and saturates at $M(T)/T \approx 1.1$ at high temperatures.

\section{Matching conditions and the collision term}

The energy-momentum tensor can be decomposed in terms of the 4-velocity $u^{\mu}$ as follows
\begin{equation}
\begin{aligned}
\label{eq:decompos-tmunu}
    T^{\mu \nu} &= \varepsilon u^{\mu} u^{\nu} - P \Delta^{\mu \nu} + h^{\mu} u^{\nu} + h^{\nu} u^{\mu} + \pi^{\mu \nu},
\end{aligned}
\end{equation}
where $\varepsilon$ is the total energy density, $P$ is the total isotropic pressure, $h^{\mu}$ is the energy diffusion, $\pi^{\mu \nu}$ is the shear-stress tensor, and we defined the projection operator $\Delta^{\mu \nu} = g^{\mu \nu} - u^{\mu}u^{\nu}$. They are obtained from moments of $f_{\bf p}$ as explained in \cite{Denicol:2021}. In general, $\varepsilon_{0}$ and $P_{0}$ may have non-equilibrium corrections, such that $\varepsilon = \varepsilon_{0} + \delta \varepsilon$, $P = P_{0} + \Pi$, respectively.

The meaning of $u^{\mu}$ and $\beta$ for non-equilibrium states is determined by matching conditions \cite{Denicol:2021}. The most widely used prescription is the one introduced by Landau \cite{landau:59fluid}, where $\delta \varepsilon \equiv 0$ and $h^{\mu} \equiv 0$. In the present work, we choose a new prescription in order to simplify Eq.~\eqref{eq:dB}. Specifically, we impose  
\begin{equation}
\label{eq:matching_kinetic1}
    \left\langle  1 \right\rangle \equiv \left\langle  1 \right\rangle_{0}, 
\end{equation}
where $\left\langle  \cdots \right\rangle_{0} \equiv \int dP \cdots f_{0\pp}$, which defines the temperature for non-equilibrium states. In this matching, $\delta \varepsilon \neq 0$. To define the 4-velocity, a further condition is needed. However, since we only consider a fluid at vanishing chemical potential, our results will not depend on this particular choice. 

With prescription \eqref{eq:matching_kinetic1},
the interaction energy can be determined solely as a function of $T$ and Eq.~\eqref{eq:dB} can be solved as if the system were in equilibrium,
\begin{equation}
\label{eq:dBeq}
\frac{\partial B(T)}{\partial T} = - \frac{g T M^2}{2\pi^2}K_1\left( \frac{M(T)}{T} \right)\frac{\partial M(T)}{\partial T},
\end{equation}
which can be readily integrated since $M(T)$ is known, and the boundary condition $B(0) = 0$ is given. Above, $K_1$ is the first modified Bessel function of the second kind. 

\subsection*{Novel relaxation time approximation}

In contrast to the traditional RTA \cite{andersonRTA:74}, in the new prescription proposed in Ref.~\cite{Rocha:2021zcw} the conservation laws hold at the microscopic level even when considering momentum-dependent relaxation times and arbitrary matching conditions. In practice, we approximate the collision term as \cite{Rocha:2022fqz}
\begin{equation}
\label{eq:nRTA}
\hspace{-1.5cm}
\begin{aligned}
C[f_{\mathbf{p}}] \approx - \frac{E_{\mathbf{p}}}{\tau_{R}} f_{0\textbf{p}} \left[ \phi_{\mathbf{p}} - 
\frac{\lan \phi_{\mathbf{p}}  \frac{E_{\mathbf{p}}^{2}}{\tau_{R}} \ran_{0}}{\lan  \frac{E_{\mathbf{p}}^{3}}{\tau_{R}}\ran_{0}} E_{\bf p}
- 
\frac{\lan \phi_{\mathbf{p}}  \frac{E_{\mathbf{p}}}{\tau_{R}} p^{\langle \mu \rangle} \ran_{0}}{ \frac{1}{3}\lan \delpp  \frac{E_{\mathbf{p}}}{\tau_{R}}\ran_{0}} p_{\langle \mu \rangle}   \right],
\end{aligned}
\end{equation}
where $\phi_{\bf p} \equiv (f_{\bf p} - f_{0\bf p})/f_{0\bf p}$. We parametrize the energy dependence of the relaxation time as $\tau_{R} = t_{R} \left( E_{\mathbf{p}}/T \right)^{\gamma}$, where the parameter $\gamma$ encodes the information of the underlying microscopic interaction, and $t_{R} > 0$. For instance, it has been argued that $\gamma = 1/2$ in QCD effective theories   \cite{dusling2010radiative}. Above, we defined the space-like projection $p^{\langle \mu \rangle} = \Delta^{\mu \nu} p_{\nu}$.

\section{Transport coefficients}

In first order theories, equation \eqref{eq:Tmunucons} is complemented by constitutive relations for the non-equilibrium currents ($ \delta \varepsilon$, $\Pi$, $\pi^{\mu \nu}$). In kinetic theory, they can be calculated using the Chapman-Enskog expansion \cite{Denicol:2021} which, when truncated at first order, leads to the following relativistic Navier-Stokes formulation of hydrodynamics, 
\begin{equation}
\begin{aligned}
& \delta \varepsilon = \chi \theta, 
\quad 
\Pi = - \zeta \theta,
\quad
\pi^{\mu \nu} = 2 \eta \sigma^{\mu \nu}.
\end{aligned}    
\end{equation}
Using \eqref{eq:nRTA}, the transport coefficients read \cite{Rocha:2022fqz}
\begin{equation}
\label{eq:coeffs-all}
\begin{aligned}
& \zeta =  -\frac{1}{3}  \left\langle \left(\Delta^{\mu \nu} p_{\mu} p_{\nu}\right) A_{\mathbf{p}} \frac{\tau_{R}}{E_{\mathbf{p}}} \right\rangle_{0} 
-
\left\langle \frac{\tau_{R}}{E_{\mathbf{p}}} A_{\mathbf{p}} \right\rangle_{0} \frac{I_{3,1}}{I_{1,0}},  
\\
& \chi = - \left\langle  A_{\mathbf{p}} \tau_{R} E_{\mathbf{p}} \right\rangle_{0} 
+ 
\left\langle \frac{\tau_{R}}{E_{\mathbf{p}}} A_{\mathbf{p}} \right\rangle_{0} \frac{I_{3,0}}{I_{1,0}}  , 
\quad
\eta = \frac{\beta}{15} \left\langle \left(\Delta^{\mu \nu} p_{\mu} p_{\nu}\right)^{2}\frac{\tau_{R}}{E_{\mathbf{p}}} \right\rangle_{0},
\end{aligned}
\end{equation}
where $A_{\mathbf{p}} = -  \beta c_{s}^{2} E_{\bf p}^{2} - \frac{\beta}{3} \Delta^{\lambda \sigma}p_{\lambda} p_{\sigma}
- \beta^{2} M \frac{\partial M}{\partial \beta} c_{s}^{2}$, and $c_{s}^{2} \equiv \left(\partial P_{0}/\partial \varepsilon_{0} \right) = \\  
(1/\beta)(I_{10} + I_{21})/[I_{30} + \frac{1}{2} I_{10}(\partial M^{2}/\partial \beta) \beta]$ is the speed of sound squared, which is expressed in terms of $I_{nq} =1/[(2q+1)!!] \left\langle \left( -\Delta^{\lambda \sigma}p_{\lambda} p_{\sigma} \right)^{q} E_{\mathbf{p}}^{n-2q}\right\rangle_{0}$.

\begin{figure}[!h]
    \centering
\begin{subfigure}{0.5\linewidth}
  \includegraphics[width=\linewidth]{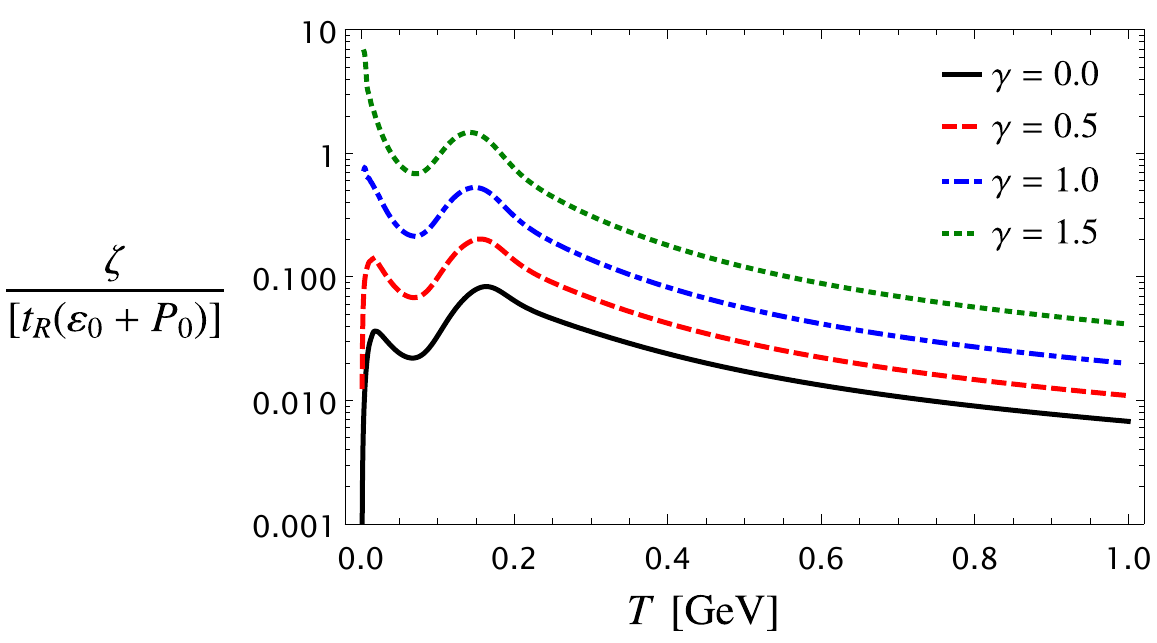}
  \label{fig:zeta-gammas}
\end{subfigure}\hfil
\begin{subfigure}{0.5\linewidth}
  \includegraphics[width=\linewidth]{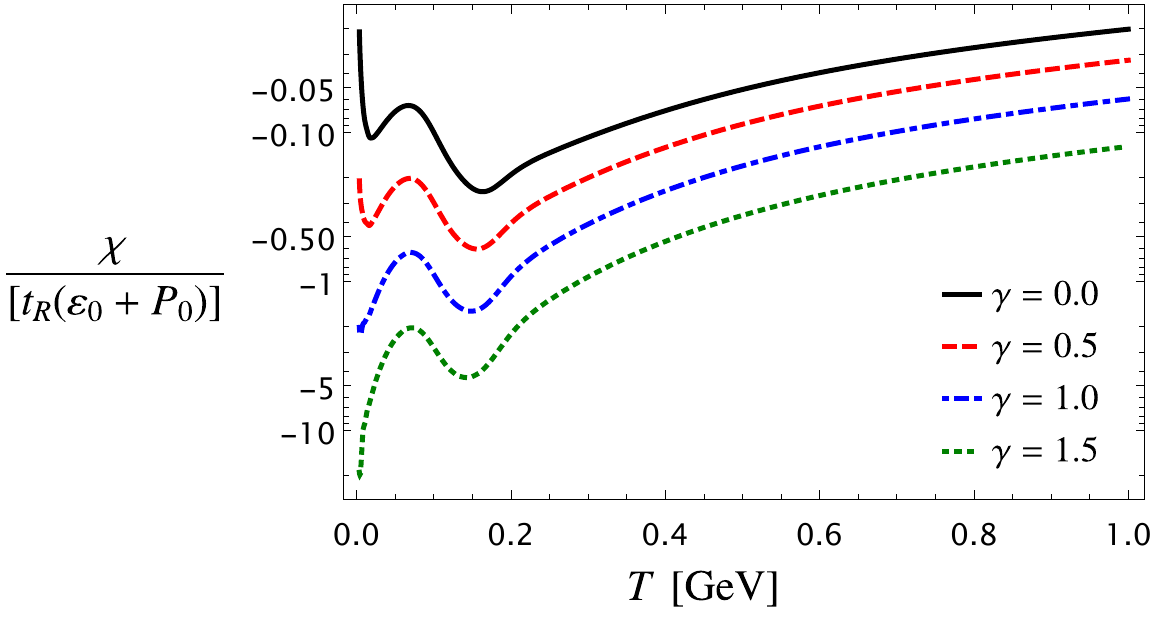}
  \label{fig:zeta-gammas}
\end{subfigure}\hfil 
\\
\begin{subfigure}{0.5\linewidth}
  \includegraphics[width=\linewidth]{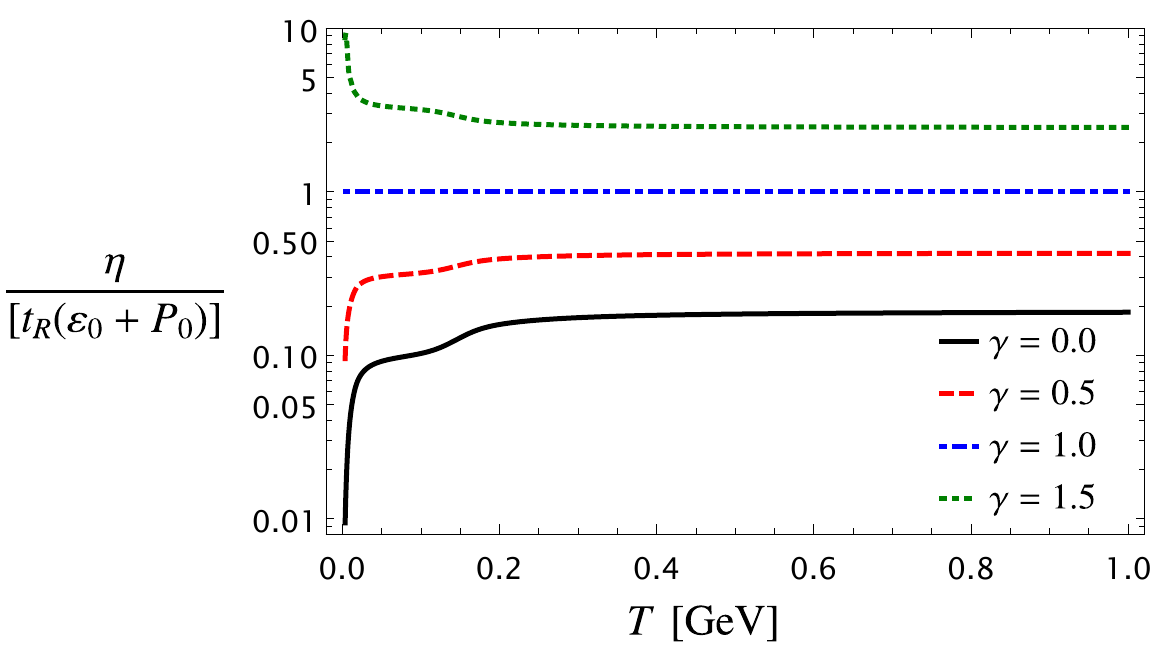}
  \label{fig:eta-gammas}
\end{subfigure}\hfil 
\begin{subfigure}{0.5\linewidth}
  \includegraphics[width=\linewidth]{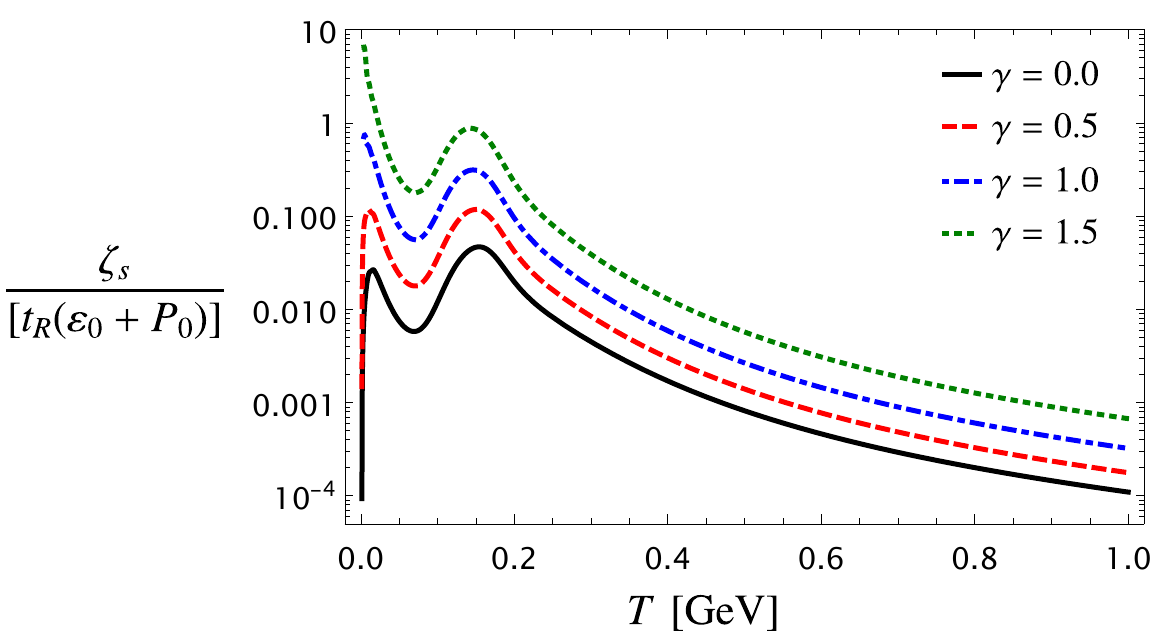}
  \label{fig:mass}
\end{subfigure}\hfil
\label{fig:deps-zeta-gammas}
\caption{(Top left) Normalized bulk viscosity, (Top right) energy correction coefficient, (Bottom left)  shear viscosity, and (Bottom right) matching-invariant bulk viscosity coefficients as functions of temperature. Each transport coefficient is shown for various values of the parameter $\gamma$.}
\label{fig:mass-coeffs}
\end{figure}

In Fig.~\ref{fig:mass-coeffs} we plot the coefficients as functions of temperature for different values of the parameter $\gamma$, as well as the temperature dependence of the mass. For all values of $\gamma$ investigated, $\zeta  \geq 0$ and  $\chi \leq 0$. In both figures, it is seen that the absolute values of the coefficients grow with $\gamma$. At low temperatures, where the effective mass is large, $M/T \rightarrow \infty$, all three normalized coefficients behave as $(M/T)^{\gamma-1}$. For $\gamma = 1$, $\eta = t_{R} (\varepsilon_{0} + P_{0})$ at all temperatures. In the opposite limit, $M/T \rightarrow 0$, $\zeta = - (1/3) \chi \propto M(T) (d/dT)\left(M(T)/T\right)$ and $\eta \sim \Gamma(\gamma + 5)/120$ \footnote{Even though this is not achieved at high temperatures, where $M/T \rightarrow 1.1$ \cite{Rocha:2022fqz}, these expansions serve as estimates.}. 
\subsection*{Entropy production}
The entropy current for classical quasiparticles is  $S^{\mu} = \int dP p^{\mu} f_{\bf p}( 1 - \ln f_{\bf p} )$. We note that the entropy production does not depend on the choice of matching conditions \cite{Rocha:2022fqz}. To first order in the Chapman-Enskog expansion, one finds
\begin{equation}
\label{eq:zeta-s}
\begin{aligned}
& 
\partial_{\mu} S^{\mu} 
\simeq 
\zeta_{s} \theta^{2}
+
2 \eta \sigma^{\mu \nu} \sigma_{\mu \nu},
\quad
\zeta_{s} =  \left\langle \frac{\tau_{R}}{E_{\bf p}} [A_{\bf p}]^{2} \right\rangle_{0} = \zeta + c_{s}^{2} \chi. 
\end{aligned}    
\end{equation}
Since both $\zeta_{s}$ and $\eta$ are non-negative, so is the entropy production. The coefficient $\zeta_{s}$ can be used to provide a matching-invariant interpretation of bulk viscosity and, indeed, for Landau matching conditions $\zeta_{s} = \zeta$. This coefficient behaves similarly to $\zeta$ as a function of temperature, with the difference that, as $M/T \rightarrow 0$, $\zeta_{s} \propto [M(T) (d/dT)\left(M(T)/T\right)]^{2}$, thus displaying a steeper descent at high temperatures in Fig.~\ref{fig:mass-coeffs}.

\section{Conclusion}

In this work we have computed the first-order transport coefficients of an effective kinetic model with temperature-dependent mass, using the new relaxation time approximation proposed in Ref.~\cite{Rocha:2021zcw}. We have used an alternative matching condition [Eq.~\eqref{eq:matching_kinetic1}] to simplify the computations, which in turn imply that there are nonzero out-of-equilibrium corrections to the energy density. We find that all transport coefficients are significantly affected by the choice of the parameter $\gamma$, which defines how the relaxation time depends on energy. Consistency with the second law of thermodynamics is demonstrated and used to derive a matching-invariant bulk viscosity coefficient. In future work, we intend to compute the transport coefficients that appear in other theories of hydrodynamics \cite{Denicol:2012cn,Bemfica:2017wps} using the present model.   

\section*{Acknowledgements}

\noindent
G.S.R. and G.S.D thank Conselho Nacional de Desenvolvimento Científico e Tecnológico (CNPq) for support. G.S.D. also thanks Fundação Carlos Chagas Filho de Amparo à Pesquisa do Estado do Rio de Janeiro (FAPERJ) process No. E-26/202.747/2018 for support. M.N.F is supported by the Funda\c c\~ao de Amparo \`a Pesquisa do Estado de S\~ao Paulo (FAPESP) grants 2017/05685-2 and 2020/12795-1. J.N. is partially supported by the U.S.~Department of Energy, Office of Science, Office for Nuclear Physics under Award No. DE-SC0021301.


\bibliographystyle{IEEEtran}
\bibliography{bibliography}

\end{document}